\documentclass[12pt,english,aps,prb,reprint,amsmath,amssymb,onecolumn,notitlepage]{revtex4-2}
\usepackage[T1]{fontenc}
\usepackage[latin9]{inputenc}
\usepackage{color}
\usepackage{babel}
\usepackage{array}
\usepackage{booktabs}
\usepackage{amsmath}
\usepackage{amssymb}
\usepackage{graphicx}
\usepackage{microtype}
\usepackage[unicode=true,pdfusetitle,
 bookmarks=true,bookmarksnumbered=true,bookmarksopen=true,bookmarksopenlevel=1,
 breaklinks=false,pdfborder={0 0 0},pdfborderstyle={},backref=false,colorlinks=true]
 {hyperref}

\makeatletter

\providecommand{\tabularnewline}{\\}

\usepackage{braket}
\usepackage{mathtools}
\usepackage{graphicx}

\usepackage{microtype}
\hypersetup{linkcolor = blue,
            urlcolor  = blue,
            citecolor = blue,
            anchorcolor = blue}

\makeatother

\begin{document}
\title{Coherence/incoherence transition temperature in molecular spin}
\author{Le Tuan Anh Ho}
\email{chmhlta@nus.edu.sg}

\affiliation{Department of Chemistry, National University of Singapore, 3 Science Drive 3 Singapore 117543}
\author{Liviu Ungur}
\email{chmlu@nus.edu.sg}

\affiliation{Department of Chemistry, National University of Singapore, 3 Science Drive 3 Singapore 117543}
\author{Liviu F. Chibotaru}
\email{liviu.chibotaru@kuleuven.be }

\affiliation{Theory of Nanomaterials Group, Katholieke Universiteit Leuven, Celestijnenlaan 200F, B-3001 Leuven, Belgium}
\date{\today}
\begin{abstract}
We examine the coherence/incoherence transition temperature of a generic molecular spin. Our results demonstrates that a molecular spin with a high coherence/incoherence transition temperature should possess a low spin number and low axiality, or high spin number and high axiality. Interestingly, the latter is better protected from the magnetic noises than the former and thus be the best candidate for a robust electron-based molecular spin qubit/qudit. The transition temperature can be further optimized if a large non-axial component of the spin Hamiltonian exists. 

\end{abstract}
\maketitle
\global\long\def\hmt{\mathcal{H}}%
\global\long\def\vt#1{\overrightarrow{#1}}%

\global\long\def\chip{\chi'}%
\global\long\def\chipp{\chi''}%

\global\long\def\tn{\mathrm{tn}}%

\section{Introduction}

Magnetic materials based on molecules have gained a lot of traction in the last several decades \citep{Sessoli1993,Gatteschi2006,Gaita-Arino2019,Chilton2022,Coronado2019,Moreno-Pineda2018,Moreno-Pineda2021,Escalera-Moreno2018}. In the area of single-molecule magnets (SMMs) \citep{Gatteschi2006,bartolome2016molecular,Aravena2018,Chilton2022}, recent observations of the hysteresis at high temperature resulting from creative chemical designs and a deep understanding of the fundamental factors influencing the magnetization relaxation has generated a lot of excitement toward a future where each molecule can be a storage bit \citep{Goodwin2017,Guo2017,Guo2018}. No less important is the rapid development of the area of molecular spin qubit where the paramagnetic molecule is proposed to be used as qubit, the elementary unit for the quantum information processing, or as qudit, a simple molecular quantum processor with more functions/qubits incorporated \citep{Leuenberger2001,Jenkins2017,Moreno-Pineda2018c,Atzori2019,Gaita-Arino2019,Carretta2021a,Chilton2022}. Recently, the realization of a Grover search algorithm on a qudit based on a set of addressable nuclear spin states within a single molecule \citep{Godfrin2017} once more demonstrates the potential of this area in particular and the molecular magnetism in general. 

In SMM, one of the key characteristics of the material is the relaxation time of the magnetization where the quantum tunneling of magnetization (QTM) plays an important role due to the nanoscale size of the molecule \citep{Gatteschi2006,bartolome2016molecular,Chilton2022,Aravena2018,Moreno-Pineda2021}. At high temperature, this QTM process can be treated in an incoherent manner \citep{Garanin1997,Leuenberger2000}. The magnetization relaxation in this regime can be considered as operating in only one single relaxation mode \citep{Blum1996,Garanin2011,Ho2016,Ho2017,Ho2022b,Ho2022c}. Meanwhile, the most important requisite for a robust molecular spin qubit is a long decoherence time of the quantum superposition within the ground doublet, which necessarily requires a large tunneling frequency and a sufficiently low operating temperature so that at least the thermal noise is suppressed \citep{Gaita-Arino2019,Aravena2018,Carretta2021a,Moreno-Pineda2018c}. Essentially, a molecular spin qubit needs to operate in the coherence limit of the quantum tunneling of magnetization where at least two relaxation modes with complex conjugate relaxation rates coexist. Assuming effect of nuclear and neighboring electronic spins are negligible, a molecular spin in principle can transit between two mentioned limits, coherent and incoherent quantum tunneling, by lowering the temperature \citep{Ho2022c,Ho2022b}. The remaining question is at which temperature this coherence/incoherence transition occurs and how to calculate it. Practically, this coherence/incoherence transition temperature plays roles as 1) the upper boundary of the temperature domain where a molecular spin qubit shows the coherence; 2) the lower boundary where the incoherent approximation of the quantum tunneling of magnetization is fully invalidated. As far as we are aware, none of the research so far approaches this problem from a theoretical point of view. 

In the previous papers, we have proposed and worked out the formula of a quantity named transition decoherence rate $\gamma_{0}$ as the boundary between the coherence and incoherence relaxation of the magnetization \citep{Ho2022b,Ho2022c}. As mentioned above, considerable attention has been usually paid to the temperature where the transition between the coherence and incoherence relaxation occurs. In principle, this transition temperature can always be found given the transition decoherence rate $\gamma_{0}$. The problem is that besides the temperature, the transition decoherence rate is also a function of other characteristics of the molecular spin system such as the spin number or anisotropy of the spin system. Consequently, the expression for the transition temperature may significantly vary from one spin Hamiltonian to another spin Hamiltonian. This variation raises additional important questions in practice: which molecular spin system will have a high coherence/incoherence transition temperature and/or be insensitive to the magnetic noise? These properties are crucial for a complex candidate to be a molecular spin qubit since they may indicate that the decoherence induced by the thermal and magnetic noise is optimized. They are also of importance for the application of the incoherent quantum tunneling approximation and accordingly the interpretation of the magnetization relaxation in single-molecule magnets. Based on the results presented in the previous works \citep{Ho2022b,Ho2022c}, we will seek the answers to all of these questions.

\section{Coherence/incoherence transition temperature}

A molecular spin system of spin number $S$ (transition metal complexes) (or total angular momentum quantum number $J$ - lanthanide-based complexes) with the following generic Hamiltonian is considered: 

\begin{multline}
\hmt=\sum_{m^{\mathrm{th}}}\left(\varepsilon_{m}+\frac{W_{m}}{2}\right)\ket{m}\bra{m}+\left(\varepsilon_{m}-\frac{W_{m}}{2}\right)\ket{m'}\bra{m'}+\sum_{m^{\mathrm{th}}}\left(\frac{\Delta_{m}}{2}\ket{m}\bra{m'}+\frac{\Delta_{m}^{*}}{2}\ket{m'}\bra{m}\right)+\sum_{n^{\mathrm{th}}}\varepsilon_{n}\ket{n}\bra{n},
\end{multline}
where $m$ ($n$) indicates the doublet $m^{\mathrm{th}}$ (or the singlet $n^{\mathrm{th}}$), $W_{m}$ is the energy bias caused by the magnetic field, and $\Delta_{m}$ is either the intrinsic or field-induced tunneling splitting gap. It is worth noting that the Hamiltonian is formulated in the localized basis \citep{Garanin2011,Ho2017,Ho2022a} and the spin system $S$ is assumed to be subject to the Redfield equation \citep{Blum1996,Garanin2011}, i.e. weakly interacting with the surrounding thermal bath. 

For the mentioned system, we have figured out in the previous works \citep{Ho2022b,Ho2022c} that the transition between coherence and incoherence relaxation occurs at a specific value of the decoherence rate $\gamma$ defined as $\gamma\equiv\left(\gamma_{11'}-\Gamma_{e}\right)/2$ where $\gamma_{11'}$ is the escape rate of the ground doublet population and $\Gamma_{e}$ is the effective relaxation rate as if the tunneling splitting gap of the ground doublet is zero. Generally speaking, $\gamma$ is temperature dependent but it is up to a specific spin system that $\gamma$ can have different types of dependence on the temperature $T$. However, if we restrict our objective only to an estimation of the temperature at which there is a transition in magnetization relaxation from incoherence to coherence, we may consider a common case when the transition rate to the nearest excited doublet/singlet at energy $U$ contributes the most to the escape rate $\gamma_{11'}$ and $\gamma_{11'}\gg\Gamma_{e}$. In this case, $\gamma$ can be \emph{roughly} approximated by: 
\begin{align}
\gamma & \overset{\mathcal{O}}{\approx}\frac{3k_{B}^{5}}{\pi\hbar^{4}\rho v^{5}}\frac{U^{3}}{\exp\left(U/T\right)-1}\left|\braket{S-1|\sum_{\alpha\beta}\tilde{D}_{\alpha\beta}S_{\alpha}S_{\beta}|S}\right|^{2}\nonumber \\
 & \overset{\mathcal{O}}{\approx}\frac{3k_{B}^{5}}{\pi\hbar^{4}\rho v^{5}}\frac{S^{4}\left(2S-1\right)^{3}D^{5}}{\exp\left[\left(2S-1\right)D/T\right]-1}.\label{eq:gamma_common_case}
\end{align}
Here we use the notation $\mathrm{LHS}\overset{\mathcal{O}}{\approx}\mathrm{RHS}$ to indicate the left-hand side (LHS) is of the order of magnitude of the right-hand side (RHS). Eq. \eqref{eq:gamma_common_case} is taken from Chapter 5 of Ref. {[}\citenum{Gatteschi2006}{]} where $S$ is the spin number, $D$ is the (second-order) axial anisotropy parameter and $U\approx\left(2S-1\right)D$. 

On the other hand, the transition decoherence rate $\gamma_{0}$ is closed related to the spin Hamiltonian of the spin system, especially the ground doublet, and given by \citep{Ho2022b,Ho2022c}:
\begin{eqnarray}
\gamma_{0} & = & \begin{cases}
\Delta_{1} & \text{for\,\,\,}W_{1}=0,\\
\frac{\left(\sqrt{\Delta_{1}^{2}-8W_{1}^{2}}+3\Delta_{1}\right)\sqrt{4W_{1}^{2}-\Delta_{1}\sqrt{\Delta_{1}^{2}-8W_{1}^{2}}+\Delta_{1}^{2}}}{8\sqrt{2}W_{1}} & \text{for\,\,\,}0\le W_{1}<\frac{\Delta_{1}}{2\sqrt{2}},\\
\frac{3}{2\sqrt{2}}\sqrt{\Delta_{1}^{2}-2W_{1}^{2}} & \text{for\,\,\,}\frac{\Delta_{1}}{2\sqrt{2}}\le W_{1}<\frac{\Delta_{1}}{\sqrt{2}},\\
0 & \text{for\,\,\,}W_{1}\ge\frac{\Delta_{1}}{\sqrt{2}}.
\end{cases}\label{eq:gamma0}
\end{eqnarray}

From Eq. \eqref{eq:gamma_common_case} and \eqref{eq:gamma0}, we can now proceed to a rough estimation of the coherence/incoherence transition temperature $T_{0}$ corresponding to the transition decoherence rate $\gamma_{0}$ in both cases: at resonance $\left(W_{1}=0\right)$ and out of resonance $\left(W_{1}\ne0\right)$.

\subsection{At resonance}

From Eq. \eqref{eq:gamma_common_case}, it is straightforward that at resonance: 
\begin{gather}
T_{0}\overset{\mathcal{O}}{\approx}\frac{U}{\ln\left[1+\frac{3k_{B}^{5}}{\pi\hbar^{4}\rho v^{5}}\frac{S^{4}}{\left(2S-1\right)^{2}}\frac{U^{5}}{\gamma_{0}}\right]}\overset{\mathcal{O}}{\approx}\frac{\left(2S-1\right)D}{\ln\left[1+\frac{3k_{B}^{5}}{\pi\hbar^{4}\rho v^{5}}\frac{S^{4}\left(2S-1\right)^{3}D^{5}}{\Delta_{1}}\right]}.\label{eq:transition temperature T_0}
\end{gather}

For an anisotropic molecular spin with spin number $S$ $\left(J\right)$, the ground doublet tunneling splitting $\Delta_{1}$ depends on its spin Hamiltonian $\hmt_{\mathrm{sp}}$. This Hamiltonian has the form $\hmt_{\mathrm{sp}}=\hmt_{0}+\delta\hmt$ where the axial component $\hmt_{0}$ commutes with operator $S_{z}$ and the non-axial component $\delta\hmt$ does not. Tunneling splitting of the spin system is caused by the latter and thus may possess various forms. In order to determine the transition temperature dependence on spin number and the anisotropy of the molecular spin system, we thus need to expand $\Delta_{1}$ from the above as a function of the anisotropy and spin number of the system. This is what we show in Table \ref{tab:T0_resonance} where some typical forms of $\delta H$ \citep{Hartmann-Boutron1996a,Gatteschi2006} and the corresponding estimated $T_{0}$ derived straightforwardly from Eq. \eqref{eq:transition temperature T_0} are listed. It is apparent from the table that the transition temperature is proportional to the axial anisotropy parameter $D$ as well as approximately the spin number $S$ when these are large since the natural logarithm in the denominator of $T_{0}$ (the fourth column of Table \ref{tab:T0_resonance}) increases slower than the one in the numerator. In the case of a small $S$ and/or $D$, the transition temperature is inversely proportional to some power of $D$ since the denominator involves the logarithm function where $\ln\left(1+x\right)\approx x$ for $x\rightarrow0$. That is to say, the case of large $S$ and $D$ or the case of small $S$ and $D$ will maximize the coherence/incoherence transition temperature. 

\begin{table}
\centering{}%
\begin{tabular*}{1\textwidth}{@{\extracolsep{\fill}}>{\centering}m{3.5cm}>{\centering}m{1.2cm}>{\centering}m{3cm}>{\centering}m{2.8cm}>{\centering}m{6cm}}
\toprule 
Origin of $\Delta_{1}$ & $S$ & $\delta\hmt$  & $\Delta_{1}$ & $T_{0}$ \tabularnewline
\midrule
\midrule 
Rhombic anisotropy & Integer & $\frac{B}{4}\left(S_{+}^{2}+S_{-}^{2}\right)$ & $\frac{8DS^{2}\left(2S\right)!}{\left(S!\right)^{2}}\left(\frac{B}{16D}\right)^{s}$ & $\frac{\left(2S-1\right)D}{\ln\left[1+\frac{3k_{B}^{5}}{8\pi\hbar^{4}\rho v^{5}}\frac{\left(S!\right)^{2}S^{2}\left(2S-1\right)^{3}}{\left(2S\right)!}\left(\frac{B}{16D}\right)^{-s}D^{4}\right]}$\tabularnewline
\midrule 
Tetragonal anisotropy & Even & $C\left(S_{+}^{4}+S_{-}^{4}\right)$ & $\frac{8DS^{2}\left(2S\right)!}{\left[\left(S/2\right)!\right]^{2}}\left(\frac{C}{16D}\right)^{s/2}$ & $\frac{\left(2S-1\right)D}{\ln\left[1+\frac{3k_{B}^{5}}{8\pi\hbar^{4}\rho v^{5}}\frac{S^{2}\left(2S-1\right)^{3}\left[\left(S/2\right)!\right]^{2}}{\left(2S\right)!}\left(\frac{C}{16D}\right)^{-s/2}D^{4}\right]}$\tabularnewline
\midrule 
Transverse field & Any & $g\mu_{\mathrm{B}}H_{x}S_{x}$ & $\frac{8DS^{2}}{\left(2S\right)!}\left(\frac{g\mu_{\mathrm{B}}H_{x}}{2D}\right)^{2s}$ & $\frac{\left(2S-1\right)D}{\ln\left[1+\frac{3k_{B}^{5}}{8\pi\hbar^{4}\rho v^{5}}\frac{S^{2}\left(2S-1\right)^{3}\left(2S\right)!}{S^{2S}}\left(\frac{g\mu_{\mathrm{B}}H_{x}}{2SD}\right)^{-2s}D^{4}\right]}$\tabularnewline
\bottomrule
\end{tabular*}\caption{Estimations of the coherence/incoherence transition temperature $T_{0}$ at resonance for some typical origins of the tunneling splitting in molecular spin systems. \label{tab:T0_resonance}}
\end{table}

\begin{figure}
\begin{centering}
\begin{tabular}{ll}
a) & b)\tabularnewline
\includegraphics[width=8.5cm]{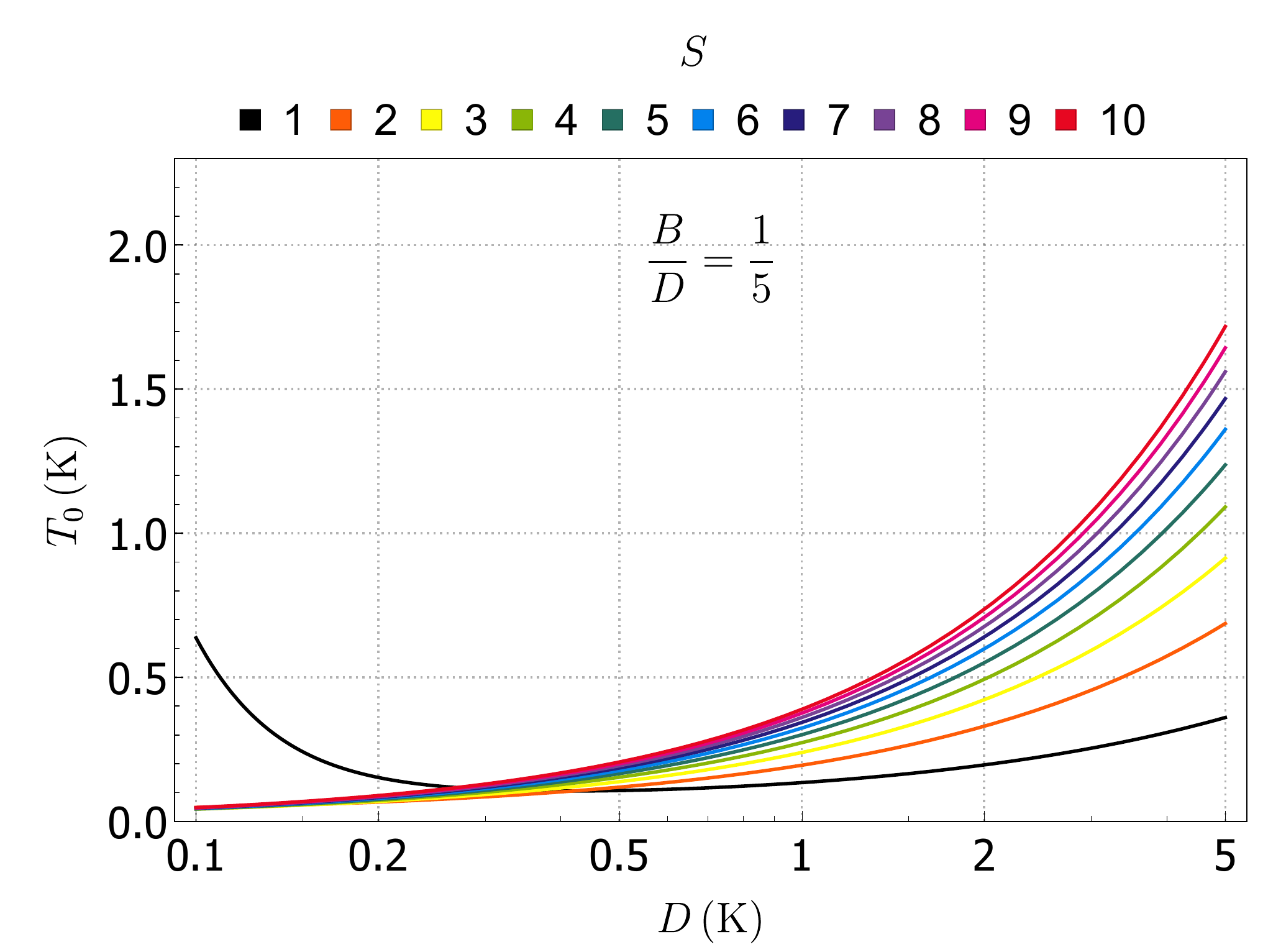} & \includegraphics[width=8.5cm]{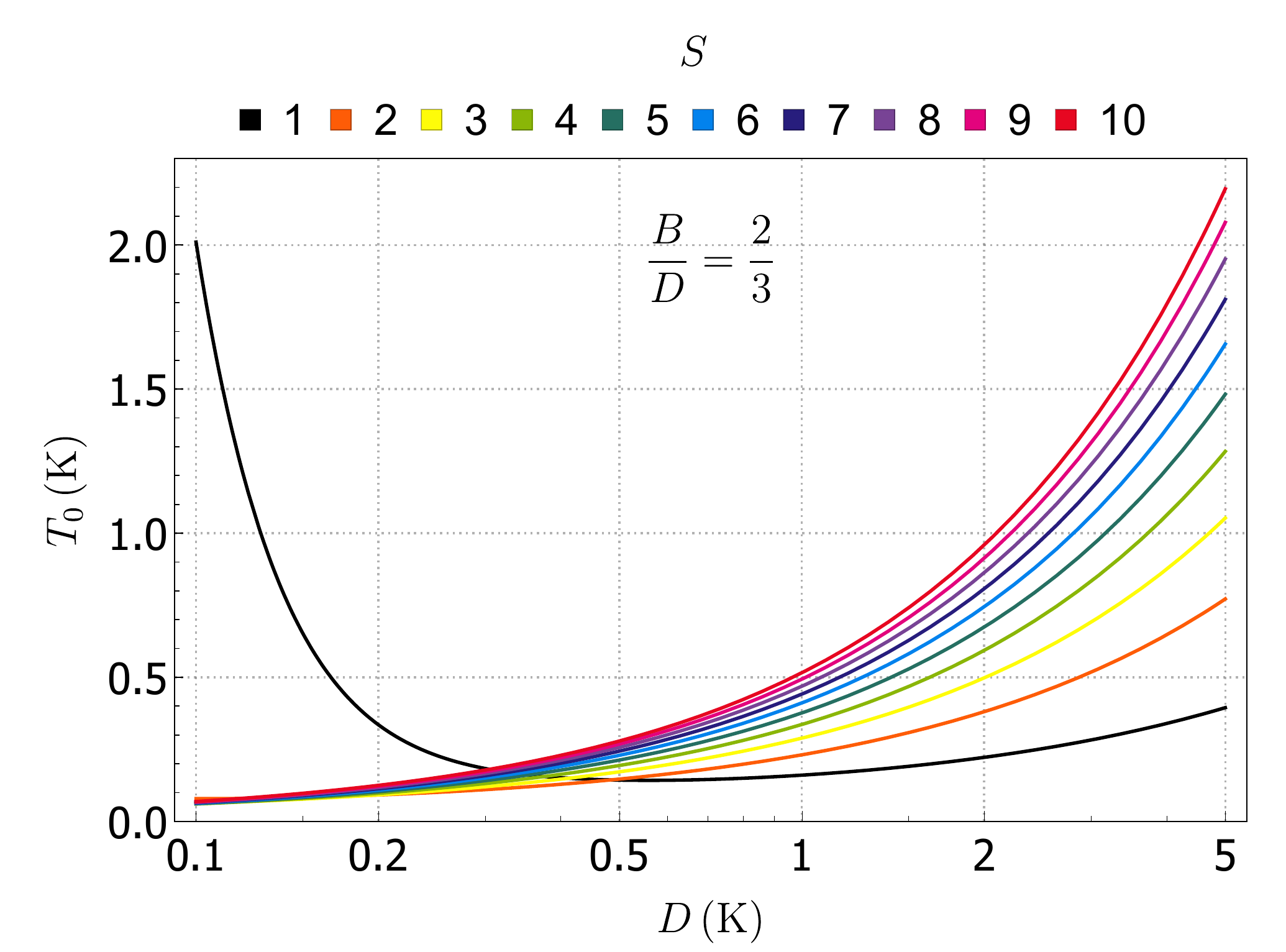}\tabularnewline
\end{tabular}
\par\end{centering}
\caption{Dependence of the coherence/incoherence transition temperature $T_{0}$ on the spin number $S$ and the axial anisotropic parameters $D$ and $B$ for a rhombic anisotropic molecular spin. \label{fig:Rhombic}}
\end{figure}

\begin{figure}
\begin{centering}
\begin{tabular}{ll}
a) & b)\tabularnewline
\includegraphics[width=8.5cm]{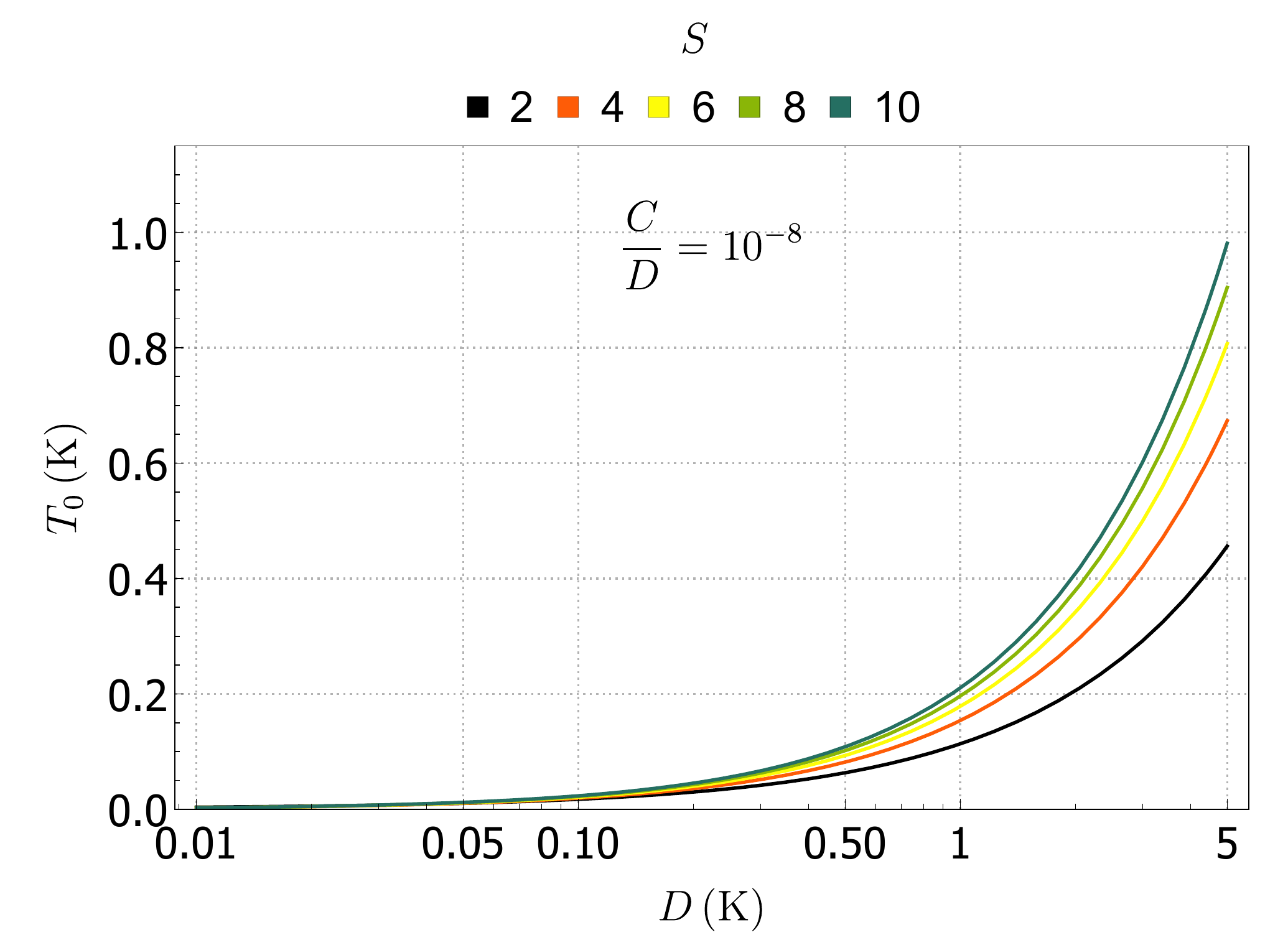} & \includegraphics[width=8.5cm]{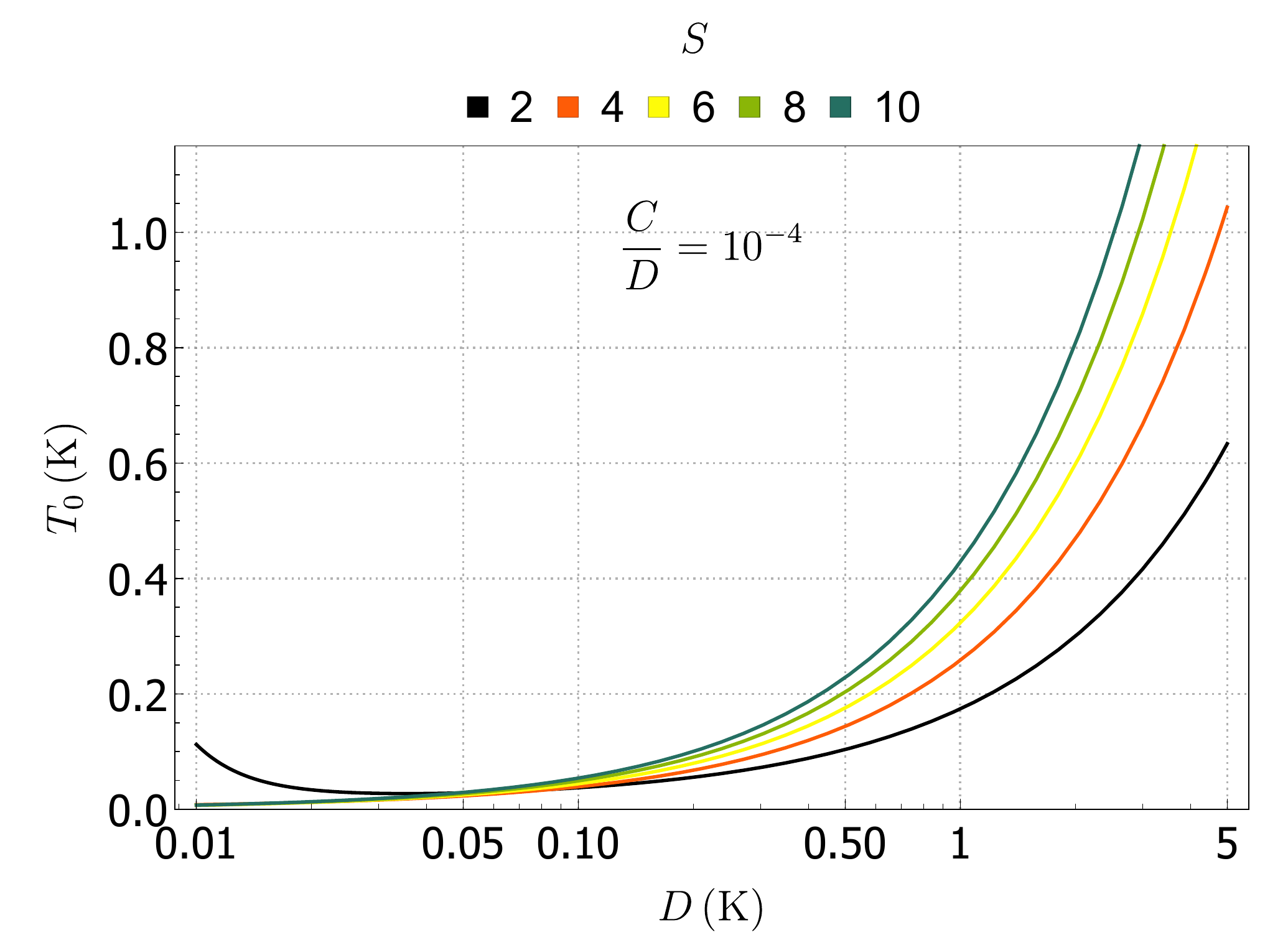}\tabularnewline
\end{tabular}
\par\end{centering}
\caption{Dependence of the coherence/incoherence transition temperature $T_{0}$ on the spin number $S$ and the anisotropic parameters $D$ and $C$ for a tetragonal anisotropic molecular spin. \label{fig:Tetragonal}}
\end{figure}

\begin{figure}
\begin{centering}
\begin{tabular}{ll}
\multicolumn{1}{l}{a)} & b)\tabularnewline
\includegraphics[width=8.5cm]{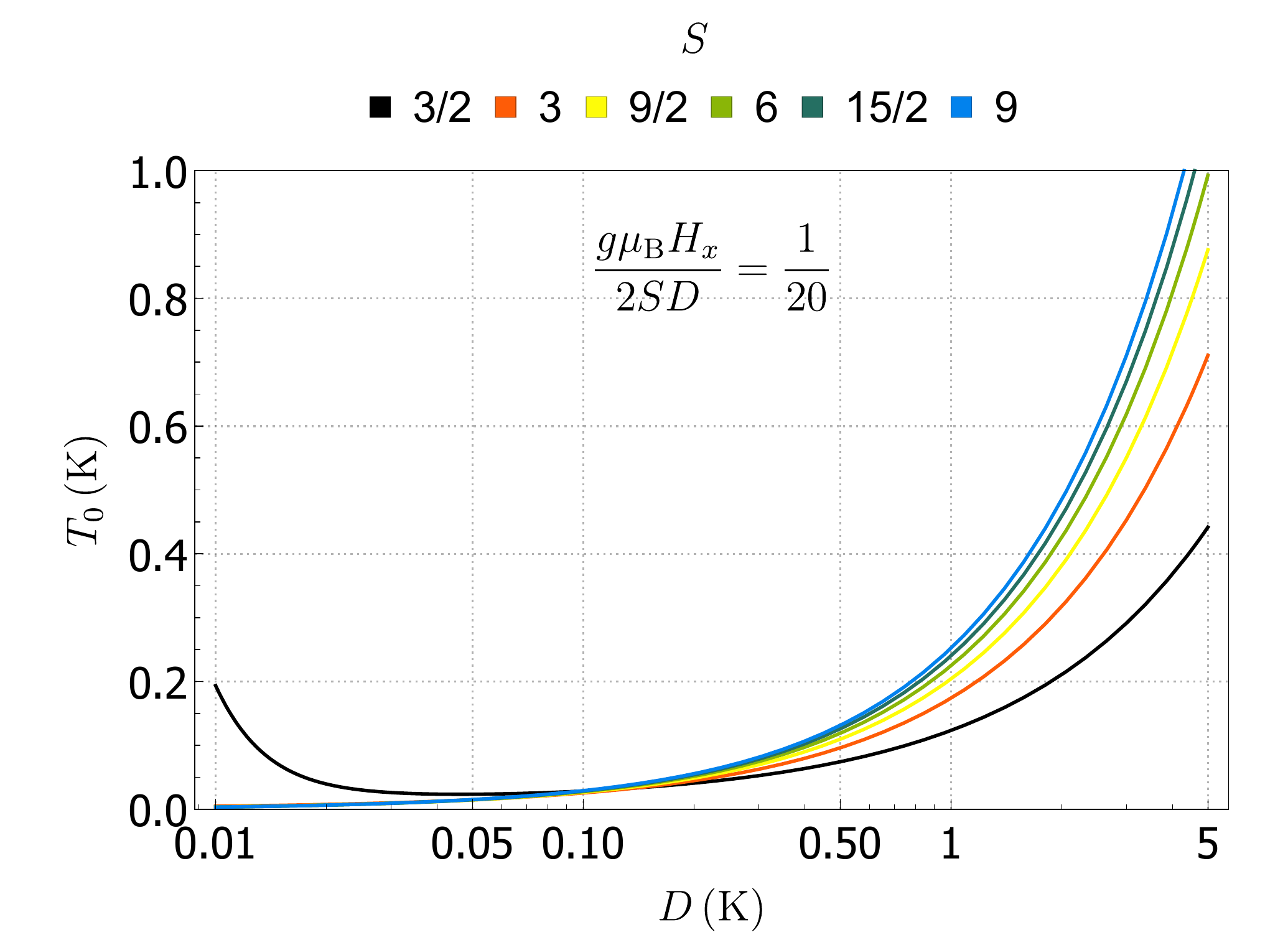} & \includegraphics[width=8.5cm]{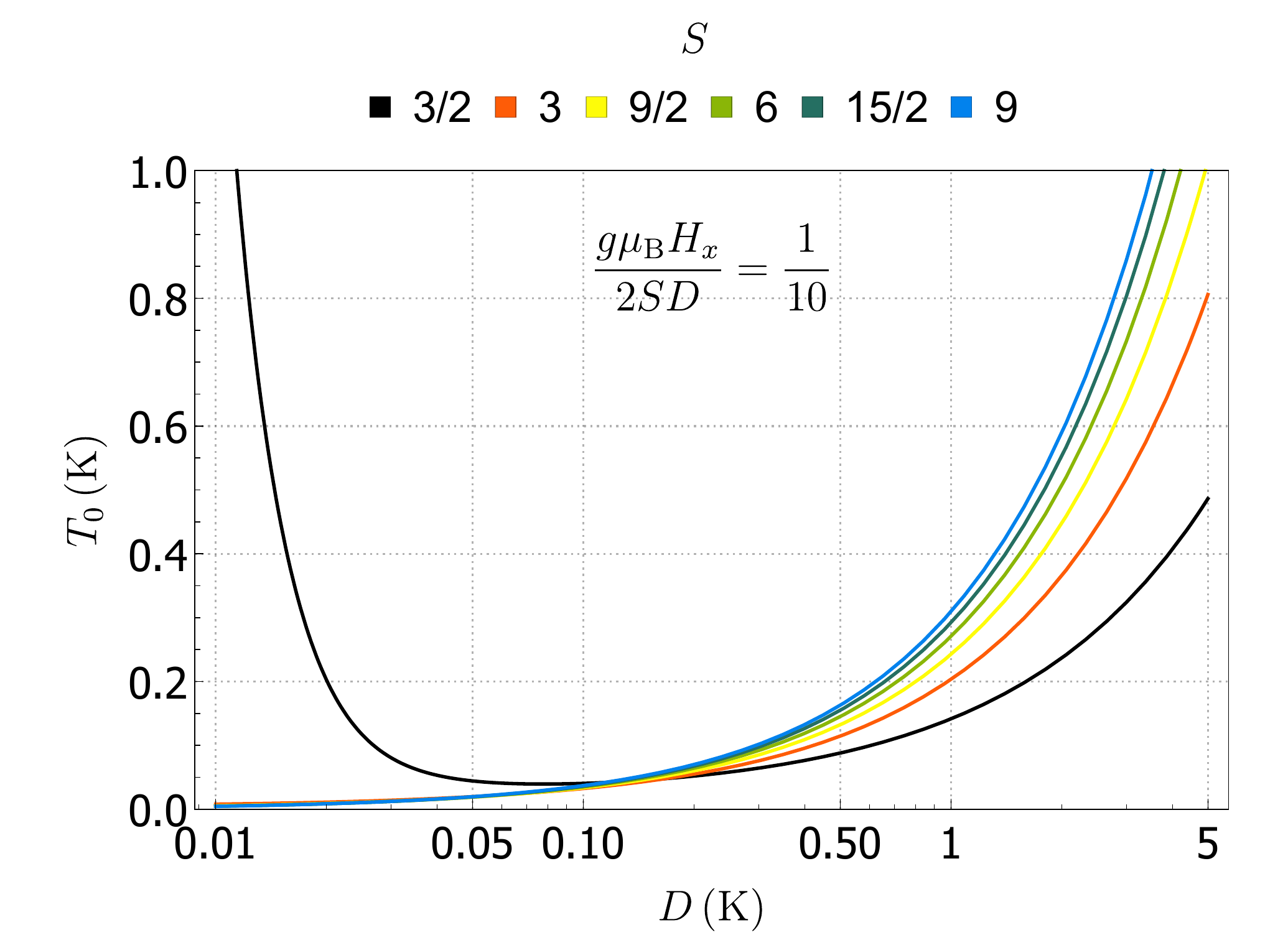}\tabularnewline
 & \tabularnewline
c) & d)\tabularnewline
\includegraphics[width=8.5cm]{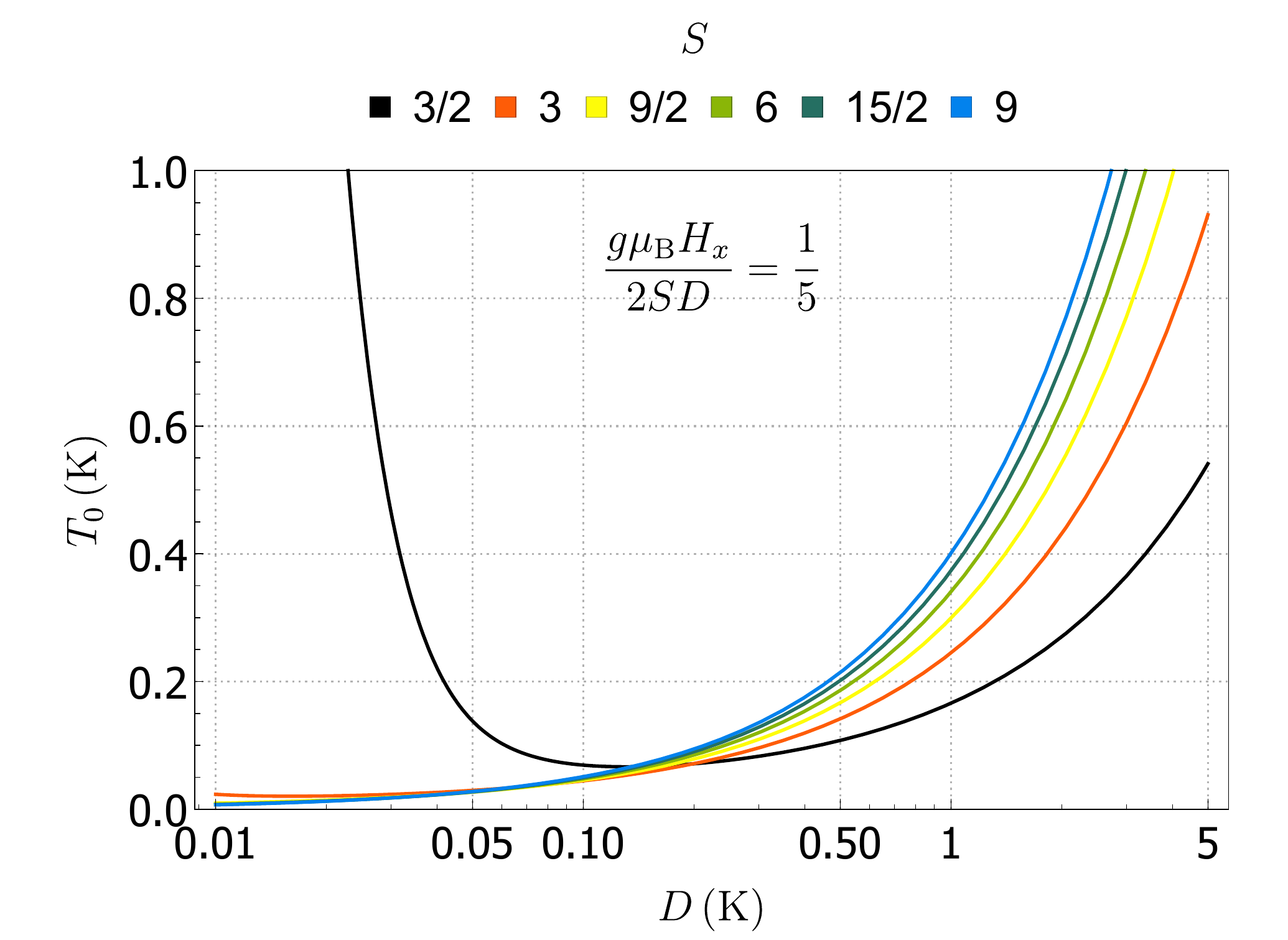} & \includegraphics[width=8.5cm]{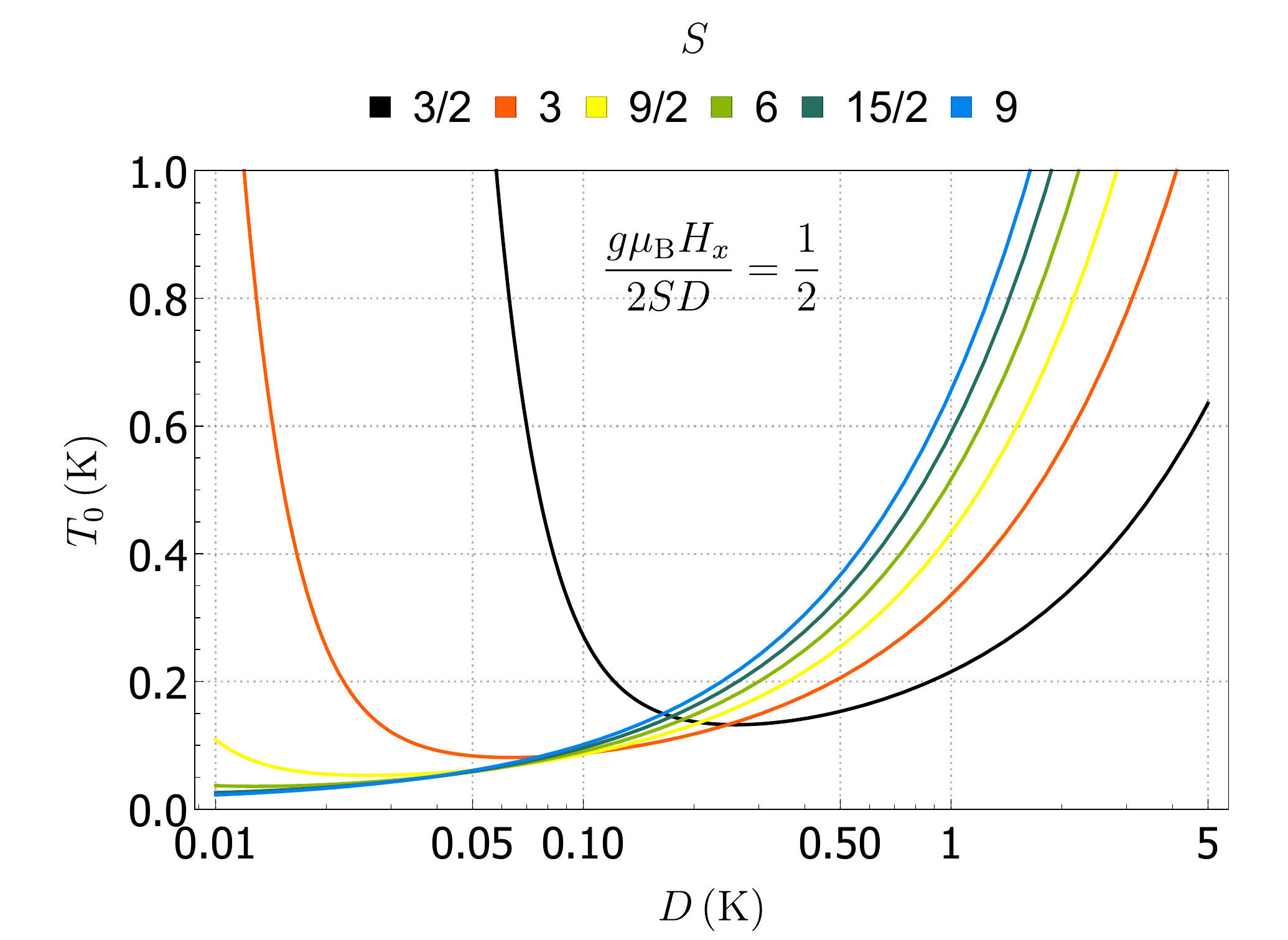}\tabularnewline
\end{tabular}
\par\end{centering}
\caption{Dependence of the coherence/incoherence transition temperature $T_{0}$ on the axial anisotropic parameter $D$ and the transversal field $H_{x}$ for an axial anisotropic molecular spin where the tunneling splitting results from a transverse magnetic field. \label{fig:transverseField}}
\end{figure}

To have a clearer view on the dependence of the transition temperature $T_{0}$ on the spin number and anisotropy of the spin system as well as giving a clue to designing a molecular spin qubit where coherence can be observed at high temperature, we illustrate this dependence in Fig. \ref{fig:Rhombic}, \ref{fig:Tetragonal}, and \ref{fig:transverseField} for difference types of the anisotropy listed in Table \ref{tab:T0_resonance}. As can be seen, regardless of the origin of the tunneling splitting, whether it is from rhombic anisotropic, tetragonal anisotropic, or transverse field, a molecular spin with a high transition temperature either must have a high spin number $S\left(J\right)$ and highly axial (large $D$), or low spin number $S$$\left(J\right)$ and low axiality (small $D$). In both cases, a large non-axial Hamiltonian component ($B$, $C$, $H_{x}$, ...) is required to maximize the transition temperature. This conclusion is highly likely applicable for any candidate for molecular spin qubit, whether it's 3d- or 4f-complexes.

The above conclusion can be explained if we consider physics of the quantum coherence in the ground doublet. In particular, for enhancing quantum coherence in molecular spin, either the ground doublet must be far from other excited states to reduce the decoherence, in particular the thermal dephasing rate, to the latter (hence high $S$ and highly axial system), or increasing the quantum tunneling splitting to amplify the effect of the quantum tunneling process (hence low $S$, low axiality). Of course, for both cases, a large non-axial Hamiltonian component always help to increase the mixing between two tunneling-split states, accordingly increase the tunneling splitting and coherence within the ground doublet.

For the well-known SMM $\mathrm{Mn_{12}ac}$, we have $C\approx5\times10^{-5}\,\mathrm{K}$, $D\approx0.5\,\mathrm{K}$, and $S=10$ (cf. Ref. \citenum{Leuenberger2000}). The transition temperature thus is estimated to be around 0.25 K at resonance. It is worth noting that throughout all calculations and figures above and hereinafter, the mass density $\rho=1.5\times10^{3}\,\mathrm{kg/m^{3}}$ and the sound velocity $v=1.5\times10^{3}\,\mathrm{m/s}$ are used.

\subsection{Out of resonance}

From $\gamma_{0}$ is given in Eq. \eqref{eq:gamma0} and the corresponding transition temperature $T_{0}$ is approximated in Eq. \eqref{eq:transition temperature T_0}, it is straightforward to obtain the function $T_{0}\left(W_{1}\right)$, 
\begin{align}
T_{0}\left(W_{1}\right) & \overset{\mathcal{O}}{\approx}\frac{\left(2S-1\right)D}{\ln\left[1+\frac{r_{0}}{r_{W_{1}}}\right]},
\end{align}
where 
\begin{eqnarray}
r_{0} & = & \frac{3k_{B}^{5}}{\pi\hbar^{4}\rho v^{5}}\frac{S^{4}\left(2S-1\right)^{3}D^{5}}{\Delta_{1}},\label{eq:r_Delta1}\\
r_{W_{1}} & = & \begin{cases}
1 & \text{for\,\,\,}W_{1}=0,\\
\frac{\left(\sqrt{1-8\frac{W_{1}^{2}}{\Delta_{1}^{2}}}+3\right)\sqrt{4\frac{W_{1}^{2}}{\Delta_{1}^{2}}-\sqrt{1-8\frac{W_{1}^{2}}{\Delta_{1}^{2}}}+1}}{8\sqrt{2}\frac{W_{1}}{\Delta_{1}}} & \text{for\,\,\,}0\le\frac{W_{1}}{\Delta_{1}}<\frac{1}{2\sqrt{2}},\\
\frac{3}{2\sqrt{2}}\sqrt{1-2\frac{W_{1}^{2}}{\Delta_{1}^{2}}} & \text{for\,\,\,}\frac{1}{2\sqrt{2}}\le\frac{W_{1}}{\Delta_{1}}<\frac{1}{\sqrt{2}},\\
0 & \text{for\,\,\,}\frac{W_{1}}{\Delta_{1}}\ge\frac{1}{\sqrt{2}}.
\end{cases}
\end{eqnarray}

\begin{figure}
\begin{centering}
\begin{tabular}{ll}
a) & b)\tabularnewline
\includegraphics[width=8.5cm]{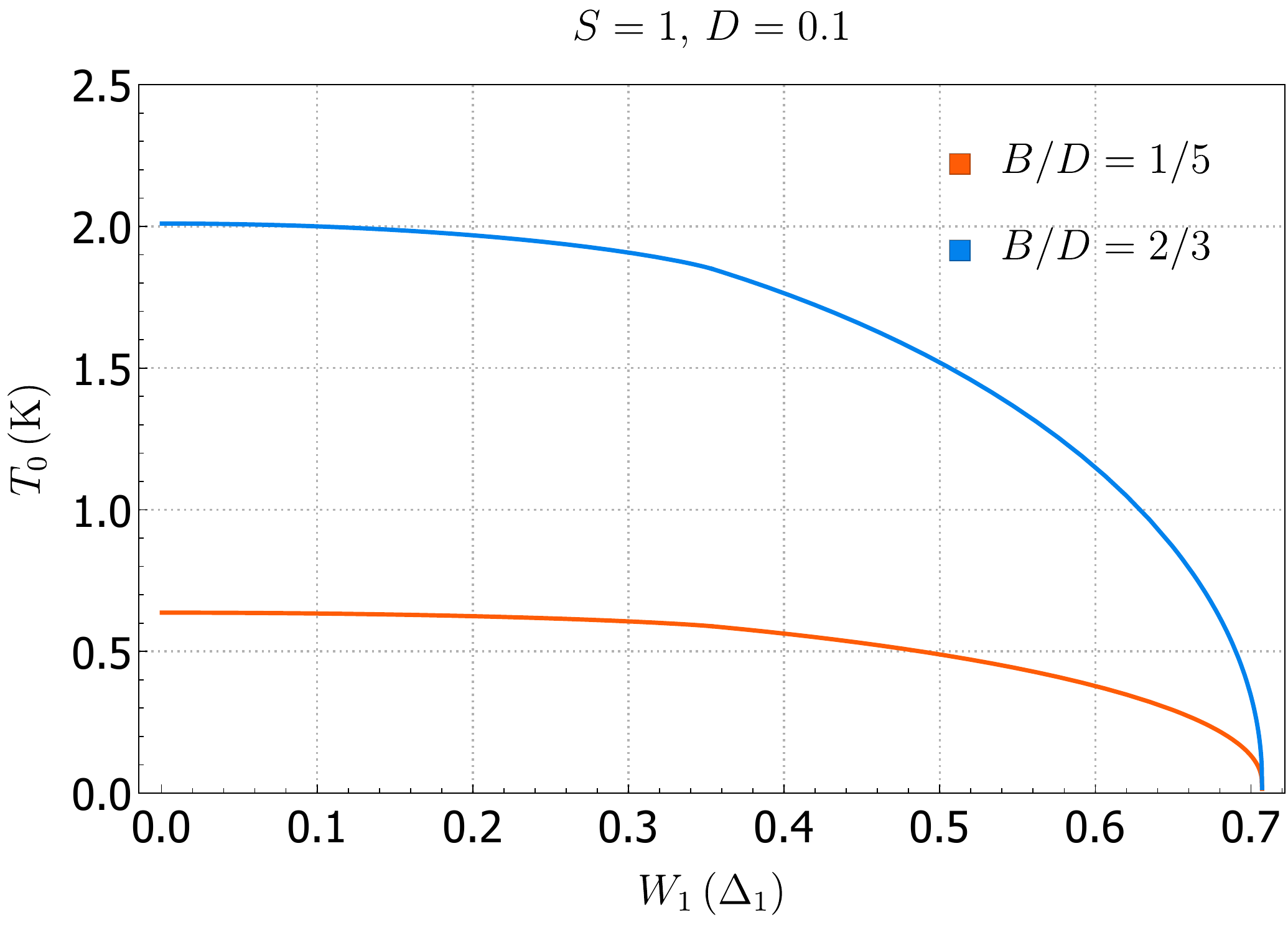} & \includegraphics[width=8.5cm]{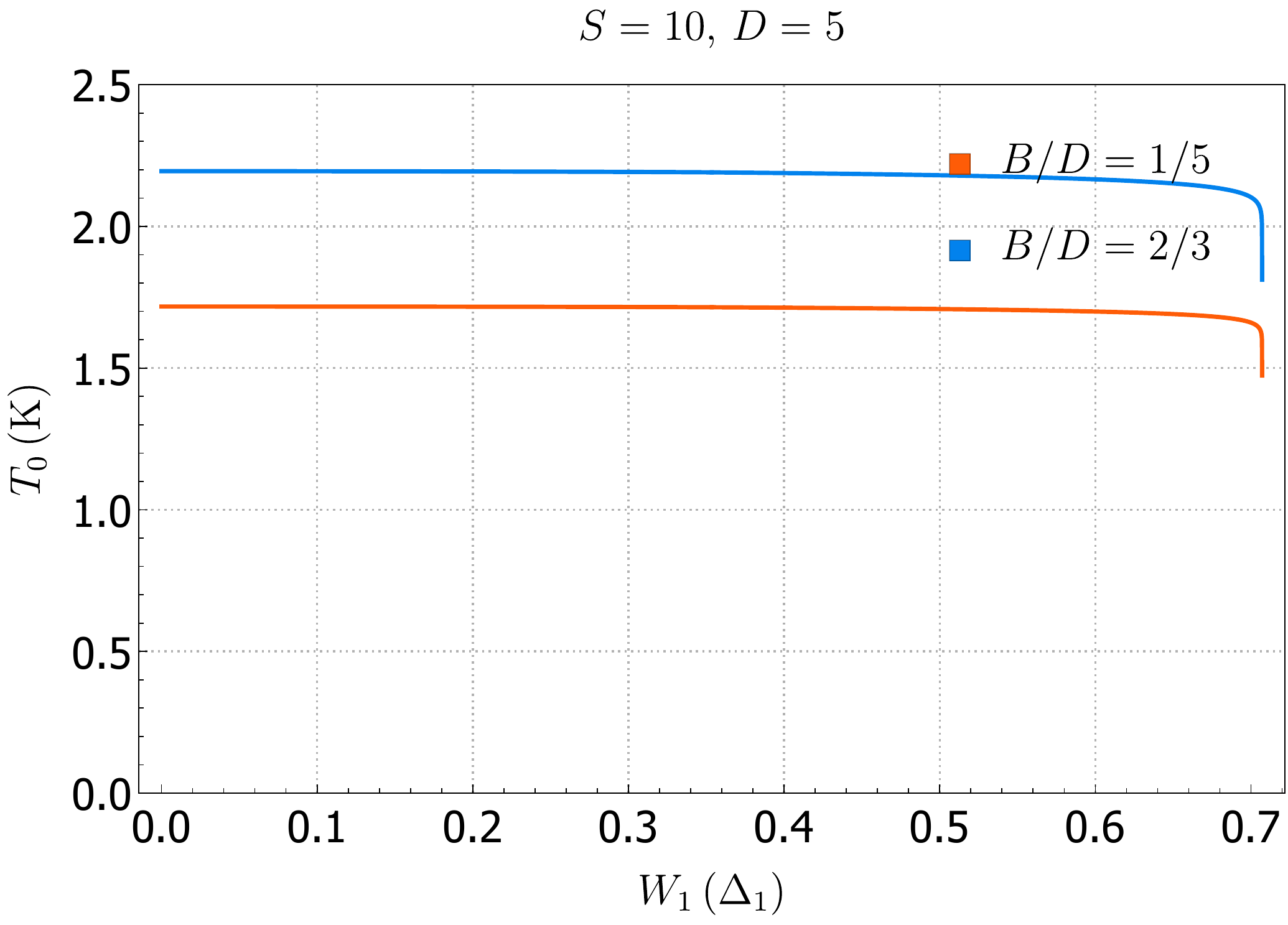}\tabularnewline
\end{tabular}
\par\end{centering}
\caption{Dependence of the transition temperature $T_{0}$ on the energy bias $W_{1}$ for a rhombic anisotropic molecular spin. \label{fig:Rhombic-W}}
\end{figure}

\begin{figure}
\begin{centering}
\begin{tabular}{ll}
a) & b)\tabularnewline
\includegraphics[width=8.5cm]{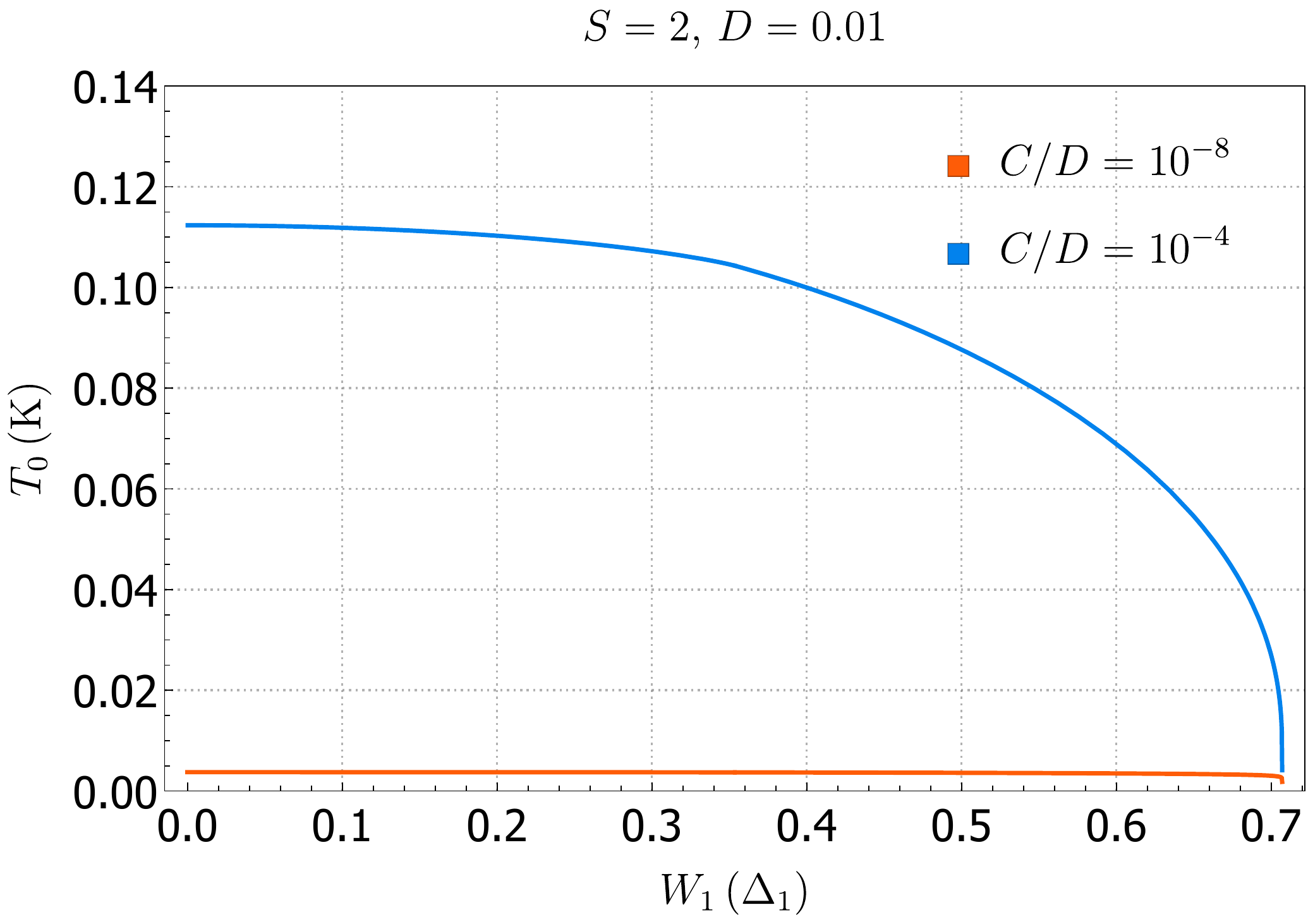} & \includegraphics[width=8.5cm]{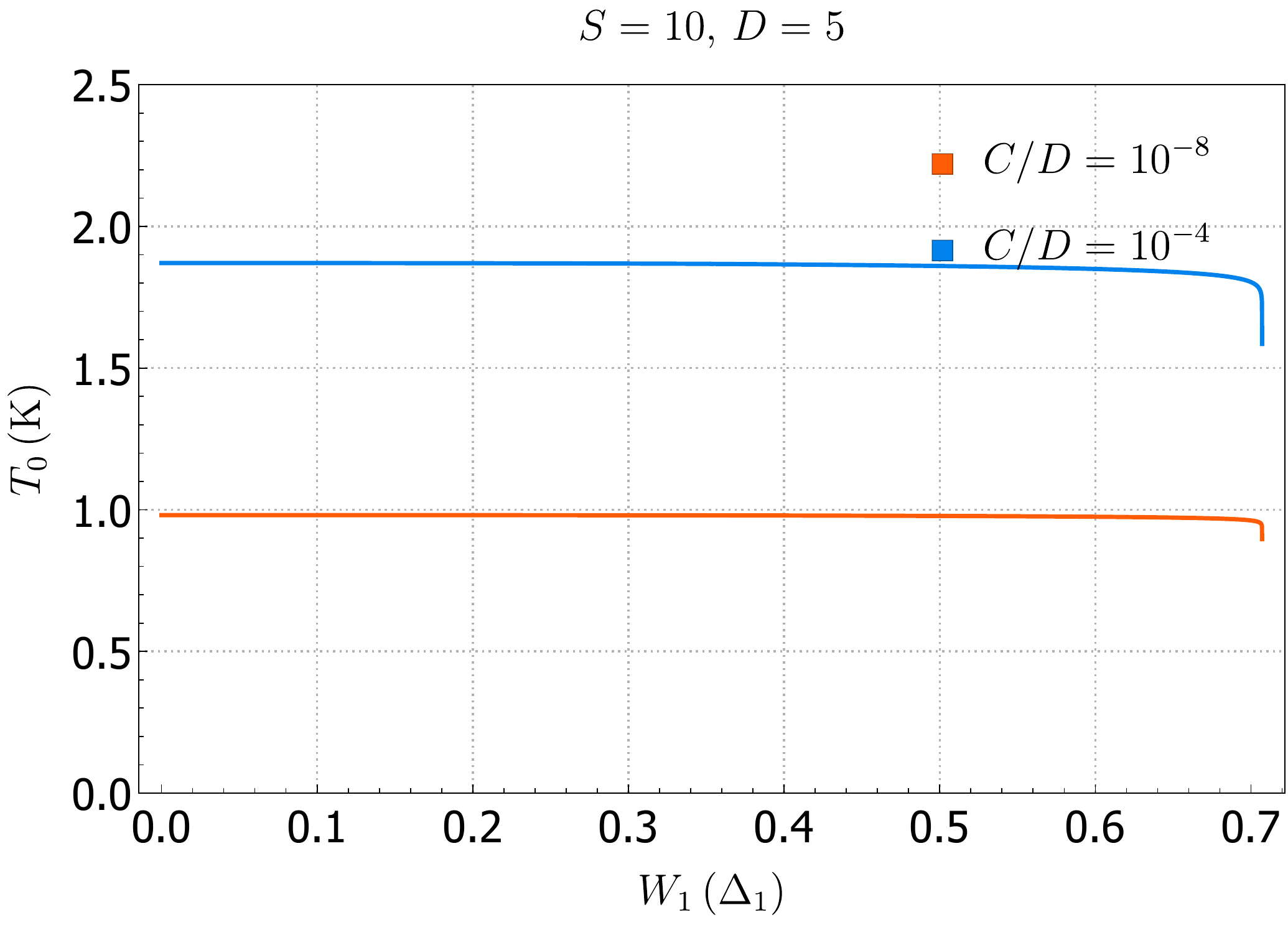}\tabularnewline
\end{tabular}
\par\end{centering}
\caption{Dependence of the transition temperature $T_{0}$ on the energy bias $W_{1}$ for a tetragonal anisotropic molecular spin. \label{fig:Tetragonal-W}}
\end{figure}

\begin{figure}
\begin{centering}
\begin{tabular}{ll}
\multicolumn{1}{l}{a)} & b)\tabularnewline
\includegraphics[width=8.5cm]{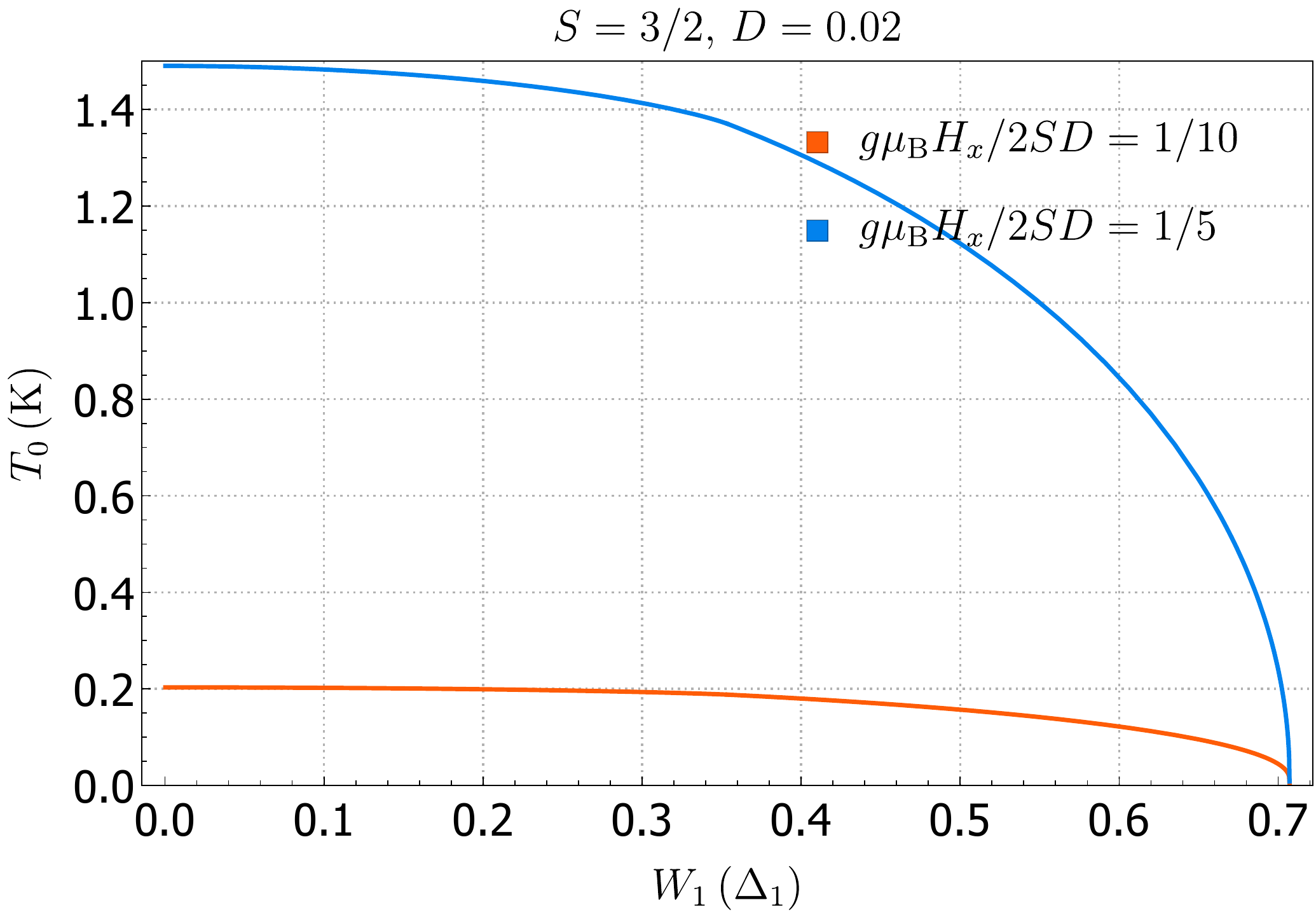} & \includegraphics[width=8.5cm]{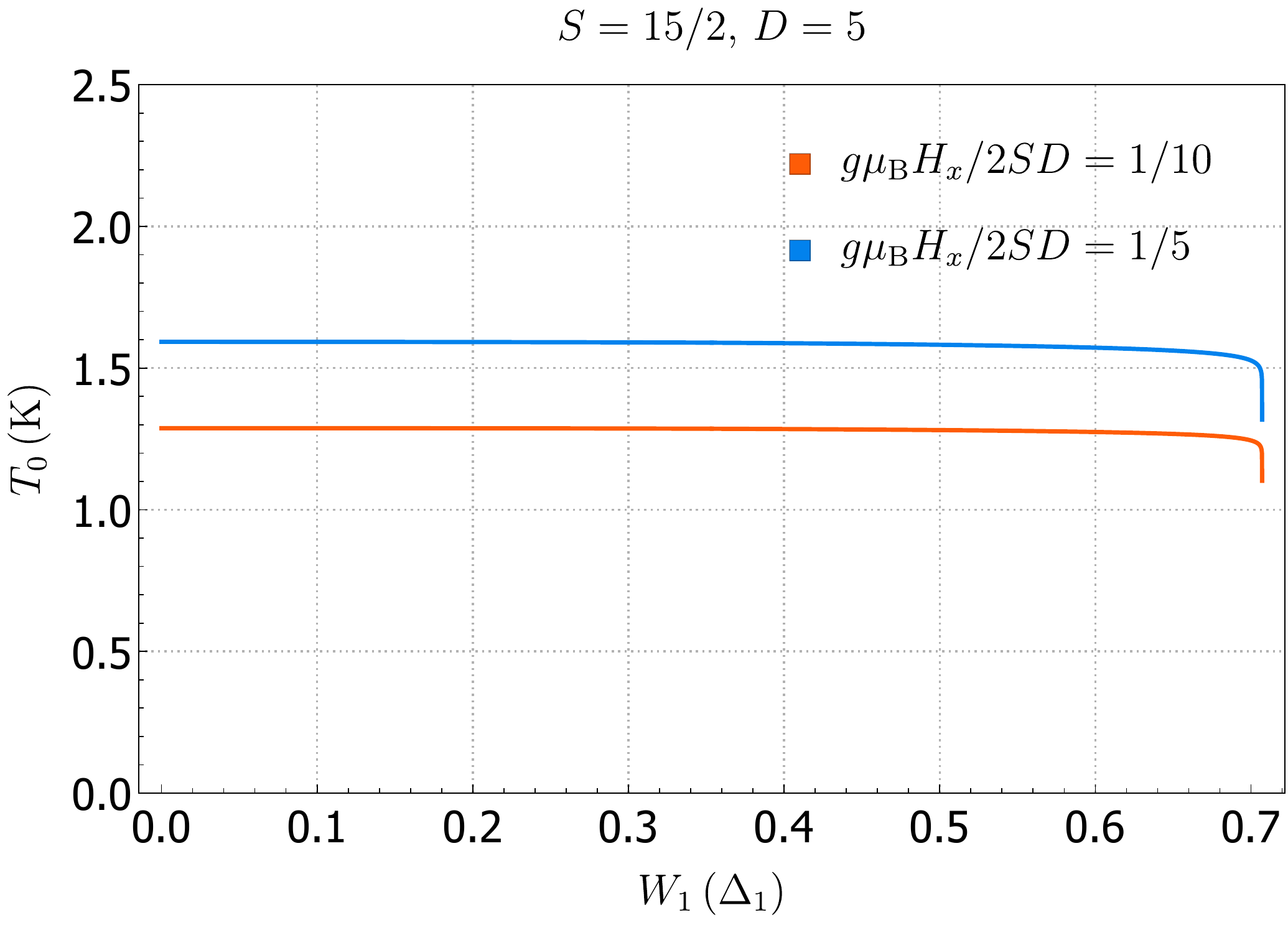}\tabularnewline
\end{tabular}
\par\end{centering}
\caption{Dependence of the transition temperature $T_{0}$ on the energy bias $W_{1}$ for an axial anisotropic molecular spin where the tunneling splitting results from a transverse magnetic field. \label{fig:transverseField-W}}
\end{figure}

For tunneling splitting $\Delta_{1}$ resulting from rhombic anisotropy, tetragonal anisotropic and transverse field, a table similar to Table \ref{tab:T0_resonance} can be easily introduced. However, we do not include it here since it is lengthy and difficult to infer any conclusion from that. Instead, we resort to a visual illustration. In Fig. \ref{fig:Rhombic-W}, Fig. \ref{fig:Tetragonal-W}, and Fig. \ref{fig:transverseField-W}, we examine the dependence of the transition temperature $T_{0}$ on the energy bias $W_{1}$ in the ground doublet for two potentially best molecular spin qubit systems concluded from the above section: 1) small spin number $S$ $\left(J\right)$ and small axial anisotropy; 2) large spin number $S$ $\left(J\right)$ with large axial anisotropy. As can be seen from these figures, the transition temperature $T_{0}$ in former system generally declines faster and faster w.r.t an increasing in the energy bias $W_{1}$ and becomes 0 at $W_{1}=\Delta_{1}/\sqrt{2}$. On the contrary, the transition temperature $T_{0}$ in the latter system with large $S$ $\left(J\right)$ and large axial anisotropy remains approximately constant until $W_{1}\approx\Delta_{1}/\sqrt{2}$. Technically, this results from a large $r_{0}$ due to a large $S$ and $D$ in the latter system but small $r_{0}$ due to a small $S$ and $D$ in the former system . This interesting dependence of the transition temperature $T_{0}$ on the energy bias $W_{1}$ suggests that those systems with large $S$ $\left(J\right)$ and highly axial anisotropy have more potential to be a practical molecular spin qubit than the one with small $S$ $\left(J\right)$ since their coherence/incoherence transition temperature, and accordingly the coherence within the ground doublet, are insensitive to the magnetic noise from the environment. This also means in order to observe the coherence, an extreme magnetic dilution can be avoided to a certain extent for this kind of molecular spin system.

\section{Discussions}

Up to now, we haven't mentioned the role of the ligand on the coherence/incoherence transition temperature. In fact, this has been implicitly included via the anisotropy parameters of the spin Hamiltonian. On the other hand, since our demonstration mainly involves a rough estimation of the transition temperature, the effect of specific properties of the ligand, such as its rigidity, on the thermal vibrations of the molecule entered the demonstration via the selection of Eq. \eqref{eq:gamma_common_case} for the transition decoherence rate $\gamma$. Regarding the static effects of other magnetic noises on the molecular spin system from surrounding nuclear or electronic spins, these are obviously covered by the energy bias $W_{1}$. However, the decoherence sources from the dynamic nature of hyperfine or dipolar interactions have not been taken into account. 

It is worth reminding that although in the main text we mainly discussed the spin number $S$, which is often a good quantum number for transition metal complexes, the results are applicable to the lanthanide-based molecular spin as well where the total angular momentum number $J$ is a good quantum number. In addition, the technique developed in this work can also be applied to other systems with different spin Hamiltonian to roughly estimate their corresponding transition temperature. 

The main objective of this paper is to seek the answer for the questions how to calculate the coherence/incoherence transition temperature and what kind of molecular spin system will have a high coherence/incoherence transition temperature and/or insensitive to the magnetic noise. Throughout an examination of the coherence/incoherence transition temperature with some typical spin Hamiltonians, it is possible to conclude that a highly axial anisotropic molecular spin with high spin number $S$ $\left(J\right)$ and a large non-axial Hamiltonian component is likely the best candidate. The second best one is a low axial spin system with low spin number $S$ but still having a large non-axial Hamiltonian component. Both kinds of system are as good at giving a high coherence/incoherence transition temperature but the former is more robust in protection against the magnetic noise. This can be explained by the energy spectrum of these systems which either disfavors the decoherence to higher energy states or amplifies the tunneling splitting gaps by which increasing the coherence within the ground doublet. To our knowledge, this is the first quantitative study to deal with these questions from the theoretical point of view and also of direct practical relevance as well. Indeed, by figuring out the characteristics of those systems with high coherence/incoherence transition temperature, the work sets out a solid guideline on which molecular spin system is the most potential to be a good molecular spin qubit. Furthermore, it also set an upper boundary for the temperature beyond which coherence in molecular spins cannot be observed as well as establishes a lower limit for the temperature below which an application of the incoherent approximation of the quantum tunneling of magnetization is no longer valid. 

\begin{acknowledgments}
L. T. A. H. and L. U. acknowledge the financial support of the research projects R-143-000-A65-133, A-8000709-00-00 and A-8000017-00-00 of the National University of Singapore. Calculations were done on the ASPIRE-1 cluster (www.nscc.sg) under the projects 11001278 and 51000267. Computational resources of the HPC-NUS are gratefully acknowledged.
\end{acknowledgments}

\bibliographystyle{apsrev4-1}
\bibliography{references}

\end{document}